\documentclass[aps,showpacs,showkeys,prd,twocolumn]{revtex4}
\usepackage{amssymb}
\usepackage{amsfonts}
\usepackage{amsmath}
\usepackage{graphicx}
\usepackage{color}
\usepackage[dvips]{epsfig}
\usepackage[dvips]{graphicx}
\usepackage{float}

\input{tcilatex}

\begin{document}

\title{Self-dual Maxwell-Chern-Simons solitons from a Lorentz-violating model%
}
\author{Rodolfo Casana and Lucas Sourrouille}
\affiliation{Departamento de F\'{\i}sica, Universidade Federal do Maranh\~{a}o,
65085-580, S\~{a}o Lu\'{\i}s, Maranh\~{a}o, Brazil.}

\begin{abstract}
Self-dual abelian Higgs system, involving both the Maxwell and Chern-Simons
terms are obtained from Carroll-Field-Jackiw theory by dimensional
reduction. Bogomol'nyi-type equations are studied from theoretical and
numerical point of view. In particular we show that the solutions of these
equations are  Nielsen-Olesen vortices with electric charge.
\end{abstract}

\keywords{Chern-Simons-like gauge theory, Topological solitons, Lorentz symmetry
violation.}
\pacs{11.10.Kk, 11.10.Lm}
\maketitle

\section{Introduction}

Lorentz and CPT violation has recently received substantial attention as a
potential signature for underlying physics, possibly arising from the Planck
scale. The most simple Lorentz- and CPT-breaking field theory consists of a
three dimensional Chern-Simons term embedded in Maxwell's four dimensional
classical electrodynamics \cite{CFJ}. Such model, usually known as
Carroll-Field-Jackiw, is part of the Standard-Model Extension (SME), the
general field-theory framework for Lorentz and CPT tests, which contains the
Standard Model of particle physics and general relativity as limiting cases
\cite{CyK}. Recent studies in the context of the CPT-odd sector of SME, have
been carried out in different areas such as radiative corrections \cite{4},
nontrivial spacetime topology \cite{5}, causality \cite{6}, supersymmetry
\cite{8}, the cosmic microwave background \cite{10}, general relativity \cite%
{11} and topological defects, specifically monopole structures was presented
in \cite{Seifert} and, the first investigation about BPS vortex solutions in
the presence of CPT-even Lorentz-violating terms of the SME was performed in
Refs. \cite{12}.

On the other hand, it is well known that the Higgs models with the Maxwell
term, support topologically stable vortex solutions \cite{NyO}. With a
specific choice of coupling constants, minimum energy vortex configurations
satisfy first order differential equations \cite{Bogo}. When this occur the
model presents another interesting feature, which says us that the model is
the bosonic sector of a supersymmetric theory \cite{LLW}.

The two dimensional matter field interacting with gauge fields whose
dynamics is governed by a Chern-Simons term support soliton solutions \cite%
{Jk}. When self-interactions are suitably chosen vortex configurations
satisfy the Bogomol'nyi-type equations with a specific sixth order potential
\cite{JW}. Another important feature of the Chern-Simons gauge field is that
inherits its dynamics from the matter fields to which it is coupled, so it
may be either relativistic \cite{JW} or non-relativistic \cite{JP}. In
addition the soliton solutions are of topological and non-topological nature
\cite{JLW}. The relativistic Chern-Simons-Higgs model described in Ref.\cite%
{JW}, was generalized to include a Maxwell term for the gauge field in Ref.%
\cite{LLK}. There, the authors, in order to maintain a notion of
self-duality, in which a lower bound for the energy is saturated by
solutions to a set of self-duality equations, introduce an additional
neutral scalar field.

In this paper we are interested in studying BPS vortices in a context of a
CPT-odd and Lorentz-violating abelian Higgs system, involving both the
Maxwell and Chern-Simons terms, which is obtained from Carroll-Field-Jackiw
theory by dimensional reduction. In particular, we will show that this model
support self-dual vortex solutions, which are different from those obtained
in Ref. \cite{LLK}. The difference lies in the fact that in our model the
neutral scalar field appears naturally from the dimensional reduction, so
that the model is self-dual without the requirement to introduce an additional neutral field such as in Ref. \cite{LLK}.
In addition we show that the vortex solution are identical in form to the Nielsen-Olesen vortices, although, in our model, the vortex possess electric charge.

\section{The theoretical framework}

Let us start by considering the Lorentz and CPT-violating
Maxwell-Chern-Simons theory, proposed by Carroll, Field, and Jackiw \cite%
{CFJ}
\begin{equation}
S=\int d^{4}x\left( -\frac{1}{4}F^{\mu \nu }F_{\mu \nu }+\frac{1}{4}p_{\rho
}\epsilon ^{\rho \sigma \mu \nu }A_{\sigma }F_{\mu \nu }-A^{\nu }J_{\nu
}\right)  \label{Ac1}
\end{equation}
where $p_{\alpha }$ is a four-vector couples to electromagnetic field, which
determines a preferred direction in spacetime violating Lorentz as well as
CPT symmetry, and $F_{\mu \nu }$ is the stress tensor defined by $F_{\mu \nu
}=\partial _{\mu }A_{\nu }-\partial _{\nu }A_{\mu }$. Here, the term $A^{\nu
}J_{\nu }$ represents the coupling between the gauge field and an external
current. The metric tensor is $g^{\mu \nu }=(1,-1,-1,-1)$ and $\epsilon
^{\alpha \beta \mu \nu }$ is the totally antisymmetric Levi-Civita tensor
such that $\epsilon ^{0123}=1$.

The Lagrangian (\ref{Ac1}) leads to the following equations of motion
\begin{equation}
\partial _{\mu }F^{\mu \nu }+p_{\mu }\tilde{F}^{\mu \nu }=J^{\nu }
\label{EqM}
\end{equation}
where $\tilde{F}^{\mu \nu }=\frac{1}{2}\epsilon ^{\mu \nu \alpha \beta
}F_{\alpha \beta }$ is the dual electromagnetic tensor.

In the following we consider a model composed by the gauge field in (\ref%
{Ac1}) coupled to a Higgs field
\begin{eqnarray}
S &=&\int d^{4}x\left[ -\frac{1}{4}F^{\mu \nu }F_{\mu \nu }+\frac{1}{4}%
p_{\alpha }\epsilon ^{\alpha \beta \mu \nu }A_{\beta }F_{\mu \nu }\right.
\label{Ac2} \\
&&\hspace{1cm}\left. \frac{{}}{{}}+|D_{\mu }\phi |^{2}-V(|\phi |)\right],
\notag
\end{eqnarray}
where Greek indexes run on $0,1,2,3$, $D_{\mu }=\partial _{\mu}-ieA_{\mu }$
is the covariant derivative and $V(|\phi |)$ is a self-interacting potential
to be determined below.

Specifically, we are interested in exploring the solitonic structure of the
model obtained from (\ref{Ac2}) via dimensional reduction. In particular we
are motivated by searching of a model with Maxwell-Chern-Simons self-dual
vortex solutions different from those present in Ref. \cite{LLK}. In order
to analyze the $(2+1)$-dimensional problem it is natural to consider a dimensional
reduction of the action by assuming that the fields do not depend on one of
the spatial coordinates, say $x_{3}$. Renaming $A_{3}$ as $N$, leads to an
action that can be written as  \cite{epjc1}
\begin{eqnarray}
S\!\! &=&\!\!\int \!\!d^{3}x\left[ -\frac{1}{4}F^{\mu \nu }F_{\mu \nu }+%
\frac{1}{4}p_{3}\epsilon ^{\beta \mu \nu }A_{\beta }F_{\mu \nu }+\frac{1}{2}%
(\partial _{\mu }N)\right.   \notag \\
&&\frac{{}}{{}}-\frac{1}{2}\epsilon ^{\mu \rho \sigma }p_{\mu }N\partial
_{\rho }A_{\sigma }-\frac{1}{2}\epsilon ^{\mu \rho \sigma }p_{\mu }A_{\rho
}\partial _{\sigma }N  \label{Ac3} \\[0.2cm]
&&\left. \frac{{}}{{}}+|D_{\mu }\phi |^{2}-e^{2}N^{2}|\phi |^{2}-V(|\phi |)%
\right] ,  \notag
\end{eqnarray}
where now Greek's indexes are $0,1,2$ and the coefficient $p_{3}$ play the role
of Chern-Simons's parameter. The Gauss law is
\begin{equation}
\partial _{i}F_{0i}+p_{3}F_{12}+\epsilon _{ij}p_{i}\partial _{j}N=eJ_{0}
\label{EqM4}
\end{equation}
where $J_{0}=i\left[ \phi (D_{0}\phi )^{\ast }-\phi ^{\ast }D_{0}\phi \right]
$ is the conserved matter current.
 Integrating this equation, over the entire plane, we obtain the important consequence that any
 object with charge $Q =e\int d^2 x \rho$ also carries magnetic flux $\Phi = \int B d^2 x$ \cite{Echarge}:
\begin{eqnarray}
\Phi =-\frac{1}{\kappa} Q
\end{eqnarray}
Here, we are interested in time-independent soliton solutions that ensure
the finiteness of the action (\ref{Ac3}). These are the stationary points of
the energy which for the static field configuration reads
\begin{eqnarray}
E\! &=&\!\!\int \!\!d^{3}x\left[ \frac{1}{2}F_{i0}^{2}+\frac{1}{2}%
F_{12}^{2}+|D_{0}\phi |^{2}+|D_{i}\phi |^{2}\right.   \notag \\
&&\frac{{}}{{}}+\frac{1}{2}(\partial _{0}N)^{2}+\frac{1}{2}(\partial
_{i}N)^{2}+e^{2}N^{2}|\phi |^{2}  \label{EJP} \\
&&\left. \frac{{}}{{}}+Np_{0}F_{12}+\epsilon _{ij}Np_{i}\partial
_{j}A_{0}+V(|\phi |)\right]   \notag
\end{eqnarray}
Since we are motivated by the desire to find self-dual soliton solution, we
will choose $p_{1}=p_{2}=0$. For this particular choice, the Gauss law (\ref%
{EqM4}) takes the simple form
\begin{equation}
\partial _{i}F_{0i}+p_{3}F_{12}=eJ_{0}  \label{EqM5}
\end{equation}
This is just the Gauss law of the model proposed in Ref.\cite{LLK}. After
integration by parts and using the Gauss law, the energy functional can be
rewritten as%
\begin{eqnarray}
E\! &=&\!\!\int \!\!d^{3}x\left\{ \frac{1}{2}[F_{i0}\pm \partial
_{i}N]^{2}+|D_{i}\phi |^{2}+|D_{0}\phi \mp ie\phi N|^{2}\right.   \notag \\%
[-0.2cm]
&& \\[-0.2cm]
&&\left. +\frac{1}{2}(\partial _{0}N)^{2}+\frac{1}{2}F_{12}^{2}+N(p_{0}\pm
p_{3})F_{12}+V(|\phi |)\right\}   \notag
\end{eqnarray}
Here, the form of the potential $V(\rho )$ that we choose is motivated by
the desire to find self-dual soliton solution and coincides with the
symmetry breaking potential of the Higgs model
\begin{equation}
V(|\phi |)=\frac{\lambda ^{2}}{2}(|\phi |^{2}-v^{2})^{2}\;,  \label{pot2}
\end{equation}
To proceed, we need a fundamental identity
\begin{equation}
|D_{i}\phi |^{2}=|(D_{1}\pm iD_{2})\phi |^{2}\pm eF_{12}|\phi |^{2}\pm \frac{%
1}{2}\epsilon ^{ij}\partial _{i}J_{j}  \label{iden}
\end{equation}
Using this identity and choosing $p_{0}=\mp p_{3}$ the energy may rewritten
as
\begin{eqnarray}
E &=&\int d^{3}x\left( \frac{1}{2}[F_{i0}\pm \partial _{i}N]^{2}+|D_{0}\phi
\mp ie\phi N|^{2}\right.   \notag \\
&&+|(D_{1}\pm iD_{2})\phi |^{2}+\frac{1}{2}[F_{12}\pm e(|\phi
|^{2}-v^{2})]^{2}  \label{EJP2} \\
&&\left. \pm ev^{2}F_{12}+(\frac{\lambda }{4}-\frac{e^{2}}{2})(|\phi
|^{2}-v^{2})^{2}+\frac{1}{2}(\partial _{0}N)^{2}\right) .  \notag
\end{eqnarray}
When the symmetry breaking coupling constant $\lambda $ is such that
\begin{equation*}
\lambda =2e^{2}\;,
\end{equation*}
i.e. when the self-dual point of the Abelian Higgs model is satisfied, the
energy (\ref{EJP2}) reduce to a sum of square terms which are bounded below
by a multiple of the magnitude of the magnetic flux:
\begin{equation}
E\geq ev^{2}|\Phi |
\end{equation}
Here, the magnetic flux is determined by the requirement of finite energy.
This implies that the covariant derivative must vanish asymptotically, which
fixes the behavior of the gauge field $A_{i}$. Then we have
\begin{equation}
\Phi =\int d^{2}xB=\frac{2\pi }{e}n
\end{equation}%
where $n$ is a topological invariant which takes only integer values. The
bound is saturated by the fields satisfying the Gauss law and $\partial
_{0}N=0$. This way, the first-order self-duality equations:
\begin{eqnarray}
F_{i0}\pm \partial _{i}N &=&0,  \label{bps1} \\[0.2cm]
F_{12}\pm e(|\phi |^{2}-v^{2}) &=&0,  \label{bps2} \\[0.2cm]
D_{0}\mp ie\phi N &=&0,  \label{bps3} \\[0.2cm]
(D_{1}\pm iD_{2})\phi  &=&0,  \label{bps4}
\end{eqnarray}
For the static field configuration this set of equations reduce to
\begin{eqnarray}
N&=&\mp A_0, \label{bpss1}\\[0.2cm]
F_{12}\pm e(|\phi |^{2}-v^{2}) &=&0, \label{bpss2}\\[0.2cm]
(D_{1}\pm iD_{2})\phi  &=&0, \label{bpss3}
\end{eqnarray}
This set of equations (\ref{bps1})--(\ref{bps4}) is similar to the
Bogomol'nyi equations of the Ref.\cite{LLK}. In Ref.\cite{LLK}, the magnetic field $F_{1 2}$ is related to the fields $\phi$ and $N$ in the form
\begin{equation*}
F_{12}\pm (e|\phi |^{2}-ev^{2}+p_{3}N)=0
\end{equation*}%
This is not the situation in our Eq. (\ref{bps2}), where the magnetic field only has an explicit dependence on the Higgs field. Such
difference is also presents in the potential of both models, whereas
in Ref.\cite{LLK} it takes the form
\begin{equation*}
V(|\phi |,N)=\frac{1}{2}(e|\phi |^{2}-ev^{2}+p_{3}N)^{2}\;,
\end{equation*}%
here the potential defining BPS configurations takes the simple form (\ref%
{pot2}) with coupling constant in the self-dual point value $\lambda =2e^{2}$.
Thus, our model only possess topological solitons solutions whereas the
Maxwell-Chern-Simons solitons studied in Ref.\cite{LLK} are both topological and nontopological.

Another interesting aspect of our model refers to the set of equations
(\ref{bpss1})--(\ref{bpss3}). We see that the equations (\ref{bpss2}) and
(\ref{bpss3}) are the Bogomol'nyi equations of the abelian Higgs model
\cite{Bogo}, whereas the equation (\ref{bpss1}) is decoupled from
(\ref{bpss2}) and (\ref{bpss3}) and only related the neutral scalar field
with the gauge field $A_0$. Thus, we have vortex solutions which are identical
in form to the Nielsen-Olesen vortices, with the novelty that the gauge field
$A_0$ is not required to be gauged to zero, as in the Maxwell-Higgs model.

\section{Charged vortex configurations\label{sec2}}

Specifically, we look for radially symmetric solutions using the standard
static vortex Ansatz

\begin{equation}
\phi =vg\left( r\right) e^{in\theta },~A_{\theta }=-\frac{a\left( r\right) -n%
}{er},~A_{0}=\omega (r),  \label{ansatz}
\end{equation}%
where $n$ is the winding number of the vortex configuration. The scalar
functions $a\left( r\right) ,$\ $g\left( r\right) $ and $\omega \left(
r\right) $ are regular at $r=0$:
\begin{equation}
g\left( 0\right) =0,\;a\left( 0\right) =n,~\omega ^{\prime }\left( 0\right)
=0,  \label{bc00}
\end{equation}%
where $\omega _{0}=\omega \left( 0\right) $, and satisfy appropriated
boundary conditions when $r\rightarrow \infty $:%
\begin{equation}
g\left( \infty \right) =1,\,a\left( \infty \right) =0,\,\omega \left( \infty
\right) =0.  \label{bc01}
\end{equation}%
All boundary conditions will be explicitly established in subsection \ref%
{BBCC}.

As usual, the magnetic field is expressed as
\begin{equation}
B=-\frac{a^{\prime }}{er}.  \label{cb1}
\end{equation}

So the remaining BPS equations (\ref{bps2}), (\ref{bps4}), and Gauss's law (%
\ref{EqM5}) are rewritten as%
\begin{eqnarray}
&\displaystyle{g^{\prime }=\pm \frac{ag}{r}\,,}&  \label{bq1} \\
&\displaystyle{B=-\frac{a^{\prime }}{er}=\pm ev^{2}\left( 1-g^{2}\right) \,,}%
&  \label{bq2} \\[0.05cm]
&\displaystyle{\omega ^{\prime \prime }+\frac{\omega ^{\prime }}{r}-p_{3}B
-2e^{2}v^{2}g^{2}\omega =0\,,}&  \label{bq3}
\end{eqnarray}%
where the upper sign corresponds to $n>0$ and the lower signal to $n<0$.
We observe that for fixed $p_3$, if we have the solutions for $n>0$, the
correspondent solutions for $n<0$ are attained doing $g\rightarrow g~,\
a\rightarrow -a,~\omega \rightarrow -\omega $. We can also observe that for
fixed $n$, under the change $p_{3}\rightarrow -p_{3}$, the solutions go as $%
g\rightarrow g~,\ a\rightarrow a~,~\omega \rightarrow -\omega $.

The structure of BPS equations and Gauss's law given in Eqs. (\ref{bq1})--(\ref{bq3}) allows to
affirm that the profiles of the Higgs field, vector potential and magnetic field
are exactly the same as those correspondent to the uncharged vortices of the Maxwell-Higgs model but now
our vortices are electrically charged. This affirmation is verified by the numerical analysis shown in Figs. \ref{S_BPS} and \ref{B_BPS}.

\subsection{Analysis of the boundary conditions\label{BBCC}}

We obtain the behavior of the solutions of Eqs. (\ref{bq1}-\ref{bq3}) in the
neighborhood of $r\rightarrow 0$ using power series method,
\begin{eqnarray}
g\left( r\right) &=&G_{n}r^{n}+{\mathcal{\ldots }}  \label{bc0_g1} \\[0.08in]
a\left( r\right) &=&n-\frac{e^{2}v^{2}}{2}r^{2}+{\mathcal{\ldots }}
\label{bc0_a} \\
\omega \left( r\right) &=&\omega _{0}+\frac{p_{3}ev^{2}}{4}r^{2}+{\mathcal{%
\ldots }}  \label{bc0_w}
\end{eqnarray}

The asymptotic behavior when $r\rightarrow +\infty $ is performed by setting
$g=1-\delta g$ , $a=\delta a$, $\omega =\delta \omega $, with $\delta g$, $%
\delta a$, $\delta \omega $ infinitesimal functions to be computed. After
substituting such forms in Eqs. (\ref{bq1}-\ref{bq3}), and solving the
linearized set of differential equations, we obtain
\begin{equation}
\delta g\sim r^{-1/2}e^{-\beta r}\sim \delta \omega ~,~\ \ \delta a\sim
r^{1/2}e^{-\beta r},  \label{inf_1}
\end{equation}%
where $\beta $, a positive real number, is given by
\begin{equation}
\beta =ev\sqrt{2\ }
\end{equation}%

From Eqs. (\ref{bc0_g1}), (\ref{bc0_a}) and (\ref{inf_1}), we can see that the behavior,
in $r=0$ and $r\rightarrow\infty$, of the fields $g(r),\, a(r)$ and consequently
of the magnetic field, is exactly the same that of the Maxwell-Higgs self-dual solution.

\subsection{Numerical solutions}

We now introduce the dimensionless variable $\rho =evr$\ and implement the
following changes:
\begin{eqnarray}
&\displaystyle{g\left( r\right) \rightarrow \bar{g}\left( r\right) ,\
a\left( r\right) \rightarrow \bar{a}\left( \rho \right) ,~\omega \left(
r\right) \rightarrow v\bar{\omega}\left( \rho \right) ,}&  \notag \\[0.2cm]
&\displaystyle{p}_{{3}}\rightarrow ev{\kappa ,~B\rightarrow ev^{2}\bar{B}%
\left( \rho \right) ,~~\mathcal{E}\rightarrow v^{2}\bar{\mathcal{E}}\left(
\rho \right) .} &
\end{eqnarray}

Thereby, the equations(\ref{bq1})-(\ref{bq3}) are written in a dimensionless
form as%
\begin{eqnarray}
&\displaystyle{\bar{g}^{\prime }=\pm \frac{\bar{a}\bar{g}}{\rho }\,,}&
\label{eqt2} \\[0.15cm]
&\displaystyle{\bar{B}=-\frac{\bar{a}^{\prime }}{\rho }=\pm \left(
1-g^{2}\right) \,,}&  \label{eqt3} \\[0.15cm]
&\displaystyle{\bar{\omega}^{\prime \prime }+\frac{\bar{\omega}^{\prime }}{%
\rho }-\kappa \bar{B}-{{2\bar{g}^{2}\bar{\omega}=0}}\,.}&  \label{eqt4}
\end{eqnarray}

\begin{figure}[H]
\begin{center}
\scalebox{1}[1]{\includegraphics[width=6.75cm]{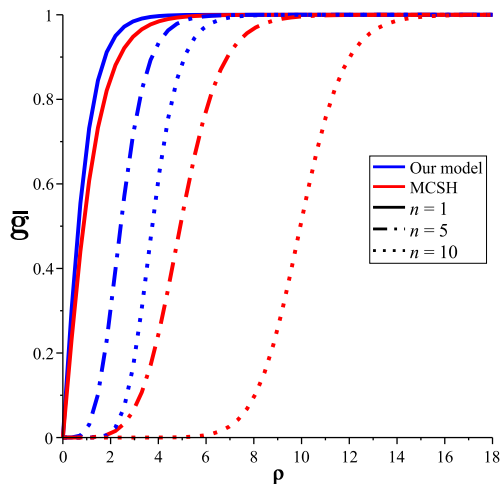}
}
\end{center}
\par
\vspace{-0.75cm}
\caption{Higgs field $\bar{g}(\protect\rho )$ (Blue lines also correspond to Maxwell-Higgs model).}
\label{S_BPS}
\end{figure}

\begin{figure}[H]
\begin{center}
\scalebox{1}[1]{\includegraphics[width=6.75cm]{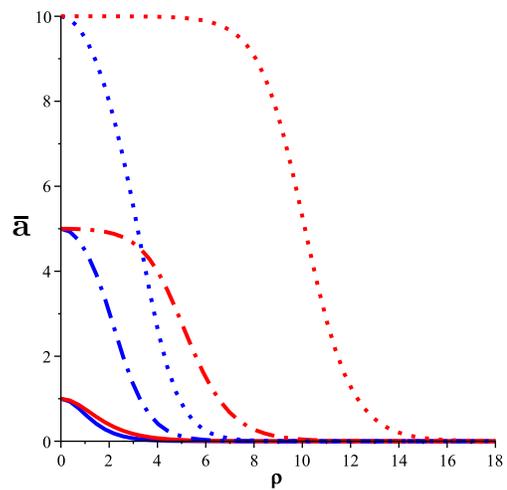}
}
\end{center}
\par
\vspace{-0.75cm}
\caption{Vector potential $\bar{a}(\protect\rho )$ (Blue lines also correspond to Maxwell-Higgs model).}
\label{A_BPS}
\end{figure}

\begin{figure}[H]
\begin{center}
\scalebox{1}[1]{\includegraphics[width=6.75cm]{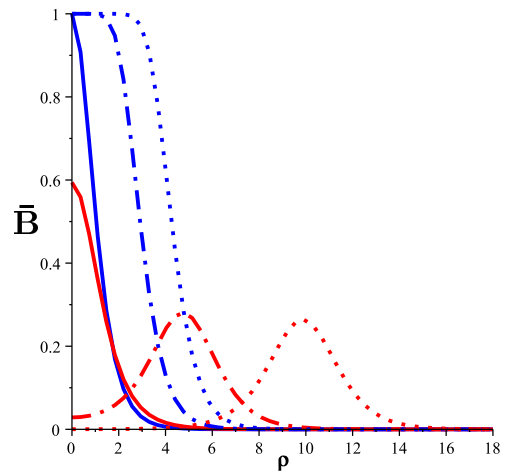}}
\end{center}
\par
\vspace{-0.5cm}
\caption{Magnetic field $\bar{B} (\protect\rho )$  (Blue lines also correspond to Maxwell-Higgs model).}
\label{B_BPS}
\end{figure}

\begin{figure}[]
\begin{center}
\scalebox{1}[1]{\includegraphics[width=7.5cm]{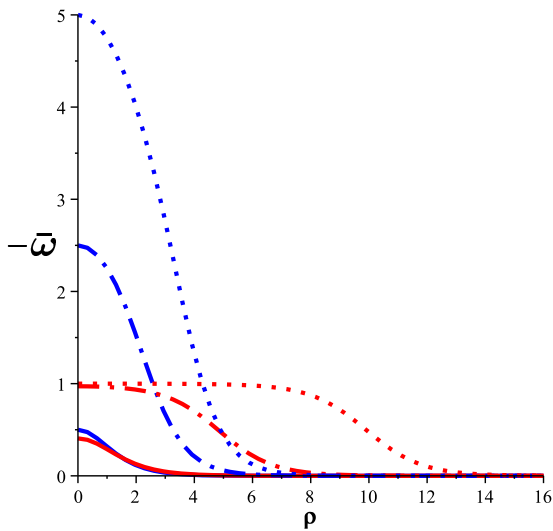}
}
\end{center}
\par
\vspace{-0.5cm}
\caption{Scalar potential $\bar{\protect\omega}(\protect\rho )$.}
\label{W_BPS}
\end{figure}

\begin{figure}[]
\begin{center}
\scalebox{1}[1]{\includegraphics[width=7.5cm]{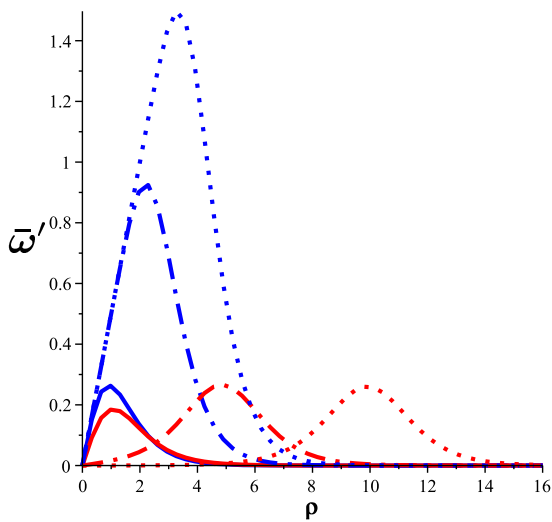}}
\end{center}
\par
\vspace{-0.5cm}
\caption{Electric field $\bar{\protect\omega}^{\prime }(\protect\rho )$.}
\label{El_BPS}
\end{figure}

We have performed the numerical analysis of the Eqs. (\ref{eqt2}--\ref{eqt4}%
) and the resultant profiles are depicted in Figs. \ref{S_BPS}--\ref{El_BPS}%
. There are shown the topological solutions with winding numbers $n=1,5, 10$
when the Chern-Simons-like parameter is fixed to be $\kappa =1$. The
profiles for our model are depicted by blue lines whereas the plots for model of Ref.\cite{LLK} (MCSH model)
are presented with red lines. The winding numbers are represented in
the following way: solid lines for $n=1 $, dash-dotted lines to $n=5$ and
dotted lines do for $n=10$. All legends are summarized in Fig. \ref{S_BPS}.

The Figs. \ref{S_BPS} and \ref{A_BPS} depict the profiles of the Higgs and
vector field, respectively. For $n=1$, the plots for both models are similar, although the difference is more evident for large
values of the winding number. In general, for fixed $n$, the profiles for
our model saturate more quickly than those from MCSH which are wider.

Fig. \ref{B_BPS} depicts the magnetic field behavior. The profiles
for $n=1$ in both models are lumps centered at origin, however the
magnetic field for the MCSH model has a smaller amplitude. On the other hand, for $n> $ 1, in
our model (blue lines), the magnetic field keeps the same amplitude in the
origin and develops a plateau around it which becomes greater as $n $ is
increased. Such behavior is very different from the magnetic field of MCSH model (red
lines) that form rings whose maximums are located away from the origin.

Fig. \ref{W_BPS} depicts the scalar potential profiles which for $\kappa>0$
always take negative values whenever $n$ positive. For $n=1$, the models
present very similar profiles, however, for $n>1$ there are notable
differences. In our model, the profiles of the scalar potential are lumps
centered in the origin whose amplitude is given by $\displaystyle{-\frac{%
n\kappa}{2}}$. These are very localized compared with the plots of MCSH model.
It is worthwhile to note that the profiles of MCSH model, for $n\gg 1$,
saturate its amplitude in the value $\displaystyle{-\frac{1}{\kappa}}$
generating a large plateau.

Fig. \ref{El_BPS} shows the electric field behavior.  For fixed $\kappa$,
our profiles develop ring structures near to the origin whose amplitude
increases with $n$  but far from the origin, the electric field
decreases to zero  satisfying the boundary conditions. This behavior
differs from the respective  MCSH  profiles which form rings, centered
long away from the origin, whose amplitude not increase  with $n$.
In general, the amplitude of the electric field are larger than those
corresponding to the MCSH  model.

\section{Conclusions and remarks}

In this paper, we have considered a self-dual system with Maxwell and Chern-Simons
terms, obtained by dimensional reduction of the Carroll-Field-Jackiw model coupled to a Higgs field.
Thus, the neutral scalar field $N$ appears as a natural consequence of the dimensional reduction
process, and the model presents self-duality equations without the necessity to introduce an additional
field in the theory.  The main difference with the self-dual model obtained in Ref. \cite{LLK} is that
in our model the magnetic vortex solution do not present explicit dependence on the neutral scalar
field. Another interesting aspect is that the vortex solutions are identical in form to the Nielsen-Olesen vortices. The only deference between the two solutions lies in the fact that our vortex solutions have electric charge.
Finally we were able to resolve numerically the Bogomol'nyi equations of the model. In addition
we analyze the results by comparing our solutions with the solutions for the self-dual system studied
in  \cite{LLK}, obtained as the main result an increase in the intensity of the magnetic and electric fields.
It is worthwhile to observe that numerical analysis shows that the profiles correspondents to the Higgs field and the magnetic field in our model are exactly the same as those correspondent to the Maxwell-Higgs model.

The study of other topological defects under effects of
Lorentz-violation are under investigations and we  expect to report on these issues in the future.

\bigskip
{\bf Acknowledgements}  \bigskip

RC thanks to CAPES, CNPq and FAPEMA (Brazilian agencies) for partial
financial support and LS thank to CAPES for full support.

\end{document}